\def\bn{\bigskip\noindent}

\def\\{$\backslash$}
\def\lsim{$_{\sim}^{<}$}
\def\hh       {{$^{h}$}}
\def\mm       {{$^{m}$}}
\def\ss       {{$^{s}$}}                  %% Overwrites TeX \ss
\def\deg      {{\ifmmode^\circ\else$^\circ$\fi} } %% Overwrites TeX \deg
\def\arcm     {{\ifmmode {'\ }\else$'     $\fi} } %% Arc minutes with  aspace
\def\arcs     {{\ifmmode{''\ }\else$''    $\fi} } %% Arc seconds with  aspace
\def\Msun     {{\ $M_{\odot}$} }

\def\nref  {\noindent\parshape 2 0.0 truein 06.5 truein 0.4 truein 06.1
truein}

\documentstyle [12pt,aaspp4]{article}

\begin{document}

\title {Radio Emission from Galaxies In The Hubble Deep Field\footnotemark[1]
\footnotetext[1]{ Based on observations with the NASA/ESA Hubble Space
Telescope, obtained
 at the Space Telescope Science Institute, which is operated by the
 Association
of Universities for Research in Astronomy, Inc., under NASA contract
NAS5-26555.}}

\author{E. A. Richards}
\affil{University of Virginia and National Radio Astronomy
Observatory\footnotemark[2]
\footnotetext[2]{The National Radio Astronomy Observatory is a facility
of the National
Science Foundation operated under cooperative agreement by Associated
Universities, Inc.}}
\authoraddr{
 520 Edgemont Road, Charlottesville, Virginia 22903 \newline
Electronic mail: er4n@virginia.edu}
\author{ K. I. Kellermann and E. B. Fomalont}
\affil{National Radio Astronomy Observatory$^2$}
\authoraddr{ 520 Edgemont Road, Charllotesville,
Virginia 22903 \newline
Electronic mail: kkellerm@nrao.edu, efomalon@nrao.edu}
\author{R. A. Windhorst}
\affil{Arizona State University}
\authoraddr{Department of Physics and Astronomy, Tempe, Arizona
85287-1504 \newline 
Electronic mail: raw@cosmos.la.asu.edu}
\and
\author{R. B. Partridge}
\affil{Haverford College}
\authoraddr{Haverford, Pennsylvania 19041 \newline
Electronic mail: bpartrid@haverford.edu}

\begin{abstract}

        We report on sensitive radio observations made with the
VLA at 8.5 GHz, centered on the Hubble Deep Field (HDF).
We collected data in the A, CnB, C, DnC, and D configurations, corresponding
to angular resolutions ranging from 0.3\arcsec \hskip5pt to 10\arcsec. 
We detected 29 radio sources in a complete sample within
4.6\arcmin \hskip5pt of the HDF center and above a flux density
limit of 9.0 $\mu$Jy (5 $\sigma$). Seven of these sources are 
located within the HDF itself, while the remaining 
22 sources are covered by the Hubble Flanking Fields (HFFs) or
ground based optical images. All of the sources in the HDF are 
identified with galaxies with a mean magnitude R = 21.7, while
the mean magnitude of the identifications outside the HDF 
is R = 22.1. Three radio sources have no optical 
counterparts to R~=~27. Based on a radio and optical
positional coincidence, we detected an additional 19 radio sources 
in this field (seven of which are contained in the HDF)
with 6.3 $\mu$Jy $\leq S_{\nu} <$ 9.0 (3.5 $\sigma \leq
S_{\nu} < 5$ ${\sigma}$) and R $\leq$ 25, but which are not
included in the complete sample.  

	The microjansky radio
sources are distributed over a wide range of redshifts (0.1 $< z <$ 3)
and have a typical monochromatic luminosity of about 10$^{23}$ W/Hz.
The majority of the optical identifications  
are with luminous (L $> L_*$) galaxies at
 modest redshifts
( 0 \lsim $z$ \lsim 1), many  with evidence for recent star-formation. The
remainder are composed of a mixture of bright field ellipticals and
late-type galaxies with evidence of nuclear
activity (LINERs and Seyferts), and nearby ($z <$ 0.2) field spirals.
None of the radio sources in this survey are
identified with quasars or galactic stars. 

\end{abstract}


\section{INTRODUCTION}

     Radio source surveys above the millijansky level are dominated
by powerful radio galaxies and quasars at substantial
redshifts. These sources are believed to be powered
by an active galactic nucleus, while a smaller fraction of these sources
are composed of lower luminosity 
starburst, Seyfert, and elliptical galaxies  at $z \simeq 0.1$.

	Below a few millijansky at 1.4 GHz, there is
an upturn in the integral radio source count, suggesting the
emergence of a different population (e.g., Windhorst {\em et al.}
1985). Many of these sub-millijansky sources have
optically blue, disk galaxy counterparts (Kron, Koo \& Windhorst
1985, Windhorst {\em et al.} 1985, Thuan \& Condon 1987).
Morphologically, these systems often appear disturbed, 
in poor groups, and/or as part of merging galaxy systems.
This evidence, coupled with the tight far infrared/radio
correlation found for many of these objects at lower
redshift (Helou {\em et al.} 1985), as well as optical
 spectra with HII-like signatures, (Benn {\em et al.} 1993)
suggests that the radio emission in these galaxies is a 
result of active star-formation. Typical star-formation rates
(SFR) for these millijansky starbursts range from 1--100 \Msun /yr
(for massive stars with $M\geq 5$\Msun). These star-formation
rates can only be maintained for short times 
($\tau$ \lsim $10^{8-9}$ per year) before depleting the available
gas in these systems. Thus, the radio emission is a sensitive measure 
of recent starburst activity.

Deeper radio surveys made with the VLA
(e.g., Fomalont {\em et al.} 1991, Windhorst {\em et al.} 1993,
1995, Kellermann {\em et al.}  1998) reach the microjansky level
and are capable of detecting powerful radio galaxies 
($P_* > 10^{25}$~W/Hz for $H_0 = 50$~km/s/Mpc and $q_0 = 0.5$),
out to $z \simeq 5$, if they exist.  However, these deep radio surveys
are dominated by starburst, Seyfert and normal elliptical and
spiral galaxies with luminosities of
$10^{21-24}$ W/Hz out to $z \simeq$ 1. Characterizing
the radio properties of field galaxies at cosmological
distances provides important clues about the AGN phenomenon
within ``normal'' galaxies, as well as those galaxies whose
bolometric luminsoity is not dominated by a central 
engine. 

	These microjansky radio surveys also provide information 
about the evolution  of rapidly evolving galaxy populations
through investigation of their star-forming activity.
Although these galaxies are typically a few hundred times brighter
at far-infrared wavelengths, the VLA microjansky radio surveys
are about 1000 times as sensitive and hence can easily 
detect luminous starbursts as distant as $z \sim 1$, 
if they exist. Moreover, high frequency $(\nu > 5$~GHz) radio
observations are unobscured by intervening dust and gas which
plague observations shortward of about 100 $\mu$m.

	Whether the microjansky 
radio sources selected at this level are the same population as the
sub-millijansky starbursts remains unclear. One indication
comes from the continuity of the submillijansky integral
source count slope, x $\sim$ 1.2, over three 
decades of flux density. The mean radio spectral index 
over this same flux density range remains constant at about
$\alpha \sim$ 0.4 (S$_{\nu} \propto \nu ^{-\alpha}$) (Fomalont
{\em et al.} 1991, Windhorst {\em et al.} 1993). 
This suggests that the source population at the millijansky
level may in fact be the same as that of the fainter,
microjansky sources. However, optical imaging and spectroscopy
by Hammer {\em et al.} (1995) for the radio sample of
Fomalont {\em et al.} (1991), complete to $S_{5 GHz}$ = 15 $\mu$Jy,
showed that many have spectra with emission lines typical of AGN
or post starburst galaxies,
reddish optical colors, and early-type galaxy counter-parts. This
evidence together with their flat ($\alpha <$ 0.5)
or inverted radio spectra ($\alpha < 0$), suggests
that the radio emission in many of these objects is
powered by low-luminosity AGN.

 	However, as shown by Condon {\em et al.} (1991) for
a nearby sample of starburst galaxies detected as radio sources,
compact radio morphology ($\theta <$ 1\arcsec) and flat radio spectra
may be explained by free-free absorption of synchrotron radiation from
nuclear starbursts, as well as by synchrotron self-absorption. Other
mechanisms which can produce a flat radio spectral index include
free-free emission from HII regions and synchrotron emission from a 
flat electron energy distribution. 

       A previous deep VLA survey complete to 7 $\mu$Jy at 8.4 GHz
(Fomalont 1996, Kellermann {\em et al.} 1998) covers an HST image made before 
refurbishment of the
Hubble telescope (Windhorst {\em et al.} 1995).  Two-thirds of the
microjansky sources in this survey are identified with galaxies having
exponential disks.  Many radio sources were also found in galaxy pairs
or
groups, suggesting that merging may be an important factor in the
radio emission mechanism. The emission lines of these objects are
consistent with those of star-forming and post-starburst galaxies at
modest redshifts ($z$ \lsim 0.7).  Fewer than 20\% of the sources in
this survey were associated with early-type galaxies or quasars.

	The radio population at the microjansky level thus appears to
be a heterogeneous mixture of low-luminosity AGN, starbursts, and
nearby spiral galaxies with typical redshifts between 0.2 and 1.
The relation (if any) between these diverse galaxy types and the
underlying radio emission mechanisms (bremsstrahlung from HII regions
and/or extended star-formation in disks, non-thermal radiation from
young supernova remnants, self-absorbed synchrotron emission from
AGNs, and/or partially absorbed non-thermal and free-free radiation
from ultra-compact nuclear starbursts) remains unclear.  Galaxy
interactions and mergers may be important.  With the high resolution of
the HST, it may be possible to distinguish amongst the many
alternatives and determine the nature of the microjansky radio emission.

    We have therefore used the VLA to make a deep radio survey at 8.5
GHz of the region containing the Hubble Deep Field and surrounding
flanking fields.  We describe the radio observations
of the HDF and the construction of the radio images in \S 2 and \S
3. In \S 4 we present the catalog of the radio sources. We discuss the
optical identifications and present images of the individual radio
sources and optical galaxies in \S 5. In \S 6 we discuss the
properties of individual sources, in particular their detections at
other wavelengths. In \S 7, we discuss the general physical properties
of the microjansky population in light of these new observations. In
\S 8 we summarize the principal results of this survey.

\section{THE RADIO OBSERVATIONS AND REDUCTIONS}
		
	We observed an area centered on the Hubble Deep
Field and located at $\alpha = 12$\hh36\mm49.4\ss~and
$\delta = 62^\circ 12\arcmin 58.00\arcsec$ (J2000) for a
total of 152 hours at 8.5 GHz.  The observations were spread among the D,
hybrid DnC, C, hybrid CnB, and A-arrays, and the observing log is given in
Table 1.

       We used the standard VLA continuum mode with four 50 MHz
channels centered on intermediate frequencies 8435 MHz and 8485
MHz observing both circular polarizations to yield an effective bandwidth of
200 MHz. The duration of each observing run was generally about eight
hours, which together with the high declination of the HDF, ensured
that $(u,v)$ coverage was uniform and symmetric, and that the number
of shadowed antennas was kept to a minimum.  The data were sampled
every 10 seconds, except for the A-configuration data which were
sampled every 3.3 seconds to decrease time-average smearing.

	In order to calibrate phase variations during the
observations, we interleaved observations of the HDF every 15 minutes
with the 0.4 Jy calibrator source 1217+585, located five degrees away.
This removed long term atmospheric phase changes as well as instrumental phase variations.
When a phase change of more than 30 degrees
between successive calibrator observations occurred, we excised the
intervening HDF data.  This discarded data in total amounted to less
than 1\% of the complete set.  We also used 1217+585 to monitor each
antenna gain over the entire observing run.  Any data associated with
an antenna whose gain varied more than 10\% between individual
calibrator observations, or which had less than 50\% of the nominal
response, was discarded. Only a few percent of the data were removed
due to these effects.

	The VLA antenna elements are
subject to systematic pointing errors as large as one arcmin due to
thermal gradients, especially at sunrise and sunset. Every three hours
we made a five-minute observation of
1302+578 to determine the antenna pointing which gave a final pointing
accuracy of about 10\arcsec.
	
    The absolute flux density scale was set by observations of 3C286
which also provided calibration for polarization measurements. The
flux density of 3C286 was assumed to be 5.17 Jy at 8435 MHz and 5.15 Jy at 8485
MHz (epoch 1995.2)(Taylor {\em et al.} 1996).  The flux density of
1217+585 was then derived from 3C286 each day and used to determine
the VLA gain change during the day.  Day to day variations in the flux
density of 1217+585 were less than 3\%, and the average flux density
measured over the entire set of observations was $0.382\pm$0.009
Jy.  This is 13\% lower than the value given by Taylor {\em et
al.}  (1996); although, between Feb 1996 to June 1997 no variations
greater than 0.02 Jy were observed in our data.

    Infrequent bursts of radio frequency interference were
isolated and removed on the basis of a 10 $\sigma$ cut over the
expected thermal {\em rms} amplitude of 15 mJy per 10 second
visibility per baseline. Also data with an amplitude deviating by more
than a factor of five from a running one-minute time averaging window
within each scan were removed.  This criteria rejected visibilities
corrupted by low level interference and cross-talk in the correlator.

    A final check on the calibration and editing of the data was
made by making separate images from each day's data.  If the expected
noise
level on the image was not reached or strange artifacts were
present, the data were reexamined and corrected.  A few reliably detected 
sources
were found to vary by about 25\% over the 18 month period.

\section{THE RADIO IMAGES}
		
     We combined all observations made from
the CnB, C, DnC, and D configurations. The relative weighting 
of each  
data point was determined from the inverse square of the noise level
at any given time, calculated from the calibrated and edited
data.  Because of the limited amount of A-configuration data (13
hours) and their very different angular resolution, these data were
not combined with the lower resolution visibilities, but imaged
separately.

     With a synthesis array such as the VLA the radio emission can be
imaged on a variety of angular scales by appropriately weighting the
visibility data.  We made images with a variety of
resolutions and a summary of their properties is shown
in Table 2.  All images were made using the NRAO AIPS package. The
dirty image was made by the gridding, convolution, and
Fourier transformation of the appropriately weighted data.  The effects
of the imperfect VLA synthesized aperture (dirty beam) were removed by
 the CLEAN
algorithm.

       All of the images cover a region much larger than the VLA primary
beam area; therefore, the detected radio sources are confined to the
central region of each image.  The rms noise listed in Table 2 was
measured in the
outer part of each image and is close to that expected from receiver
noise.  Except for the high resolution A-configuration image, the 
completeness level for each image was taken to be five times the rms noise
as is discussed in connection with the radio sample below.  There are
no peaks more negative than five times the rms noise on any of the
images.

       The image where the data points were combined with their
a priori weights (natural weighting) produces the best sensitivity for
detection of a point source.  The resolution of this image
is 3.9\arcsec \hskip5pt $\times$ 3.3\arcsec \hskip5pt (FWHM) with a
position angle on the sky of $-80^\circ$. The naturally-weighted beam
is not Gaussian-shaped, as it represents the combination of data
from the several array configurations.  After deeply cleaning
the image with 20,000 iterations, nearly all of the non-Gaussian
effects of the original beam were removed and replaced with a Gaussian
beam of 3.5\arcsec \hskip5pt resolution.  This image is shown in
Figure 1 and is referred to as the 3.5 arcsec image.

      The rms noise in the 3.5 arcsecond image is 
	  1.80 $\pm$ 0.05 $\mu$Jy outside of the
region with the radio sources; this value is close to that expected
from the system temperature, the bandwidth, and the
integration time.  As a check on the noise properties of the data, we
examined the Q and U linear polarization images.  Since the linearly
polarized emission is expected to be just a few percent of the total
intensity, the polarization image should reflect only the actual 
noise limit as it is unaffected by most
calibration errors. The rms noise in the linear polarization images
was uniform over the
image with a value of 1.72 $\mu$Jy.  The quadratic difference between
the
noise measured on the Hubble radio image of 1.80 $\mu$Jy and that of the
linear polarization images is 0.5 $\mu$Jy.  This excess noise is 
about 0.1 percent of the strongest emission in the field and it is
typical of the dynamic range limit of images made with the VLA when
there is no significant emission in the entire primary beam region.  This
extra noise is probably caused by residual atmospheric phase errors
and low level interference.

We also observed for about 13 hours in the A-configuration where
the
resolution is 0.3\arcsec.  The rms noise in the high resolution image
made from this
data alone is $6.1~\mu$Jy.
At this high resolution, five sources were detected within 2 arcmin 
of the center of the HDF above the completeness level of $37~\mu$Jy.  Since this
resolution is an
order of magnitude more than that for the other images, it not useful
to combine this data with that of the lower resolution data.

\section{THE RADIO SOURCE CATALOG}

\subsection{The Radio Source Selection}
	
    The region of sky with good sensitivity is contained in a radius
of 4.6 arcmin from the phase center.  This is the point where the sky
response of the telescope drops to 1/12 of the on axis value.
We define this as the region of completeness of the radio source
catalog.

    As seen in Table 2, the completeness of the survey is a 
function of angular resolution, and care must be taken in defining
a statistically complete radio sample. For purposes of point source
detection, the 3.5 arcsec image provides the best
sensitivity.  The noise properties of this image are Gaussian so that
in the 25,000 independent beams within 4.6 arcmin of the phase
center, less than one false source above 5 $\sigma$ is expected.
This is consistent with the most negative observed pixel value of --4.8
$\sigma$. Thus we define the
99\% completeness limit for sources to be image peak flux density,
S$_{p}$, of 9.0 $\mu$Jy for the 3.5 arcsec resolution image.

     We detected a total of 37 radio sources above the completeness
limit of $9.0~\mu$Jy in this image. Four additional, extended sources 
were detected in the 6 or 10 arcsec resolution images above 
their five sigma completeness limits.
Twelve of these 41 sources are
more than 4.6\arcm from the center of the field and are not included
as part of the complete sample.
There is no lack of confidence on the reality of
these sources; however they are so far from the center of the field
that the flux density correction due to the primary beam attenuation
is very uncertain (e.g., Windhorst {\em et al.} 1993).
Thus, the complete radio sample in the Hubble
field is composed of 29 sources with an image peak flux density 
above $9.0~\mu$Jy and within 4.6\arcmin~of the field center.

\subsection{The Source Parameters and Errors}
	
     We used the AIPS Gaussian fitting algorithm IMFIT to determine the
centroid positions, image peak and integrated flux densities of
the radio sources. The
measured image flux densities were then corrected for the primary beam
attenuation.  If the source was significantly broadened, 
we determined the major axis, minor axis and position
angles of the source.  Although the most accurate information comes
from the 3.5 arcsec resolution image, fits to the radio sources in
other images were also made.  For a few bright sources, the
0.3\arcsec~image (made from the A-configuration data) provided very
accurate positions and angular size information.

	The absolute flux density scale is tied to the flux
density of 3C286 which is known to about 3 percent accuracy.  The radio
position frame is tied to the FK5/J2000 quasar inertial reference
frame (Patniak {\em et al.} 1992) to an accuracy of about 0.02\arcsec~and is
discussed below where the radio-optical alignment is considered.
A description of other error terms in deep VLA images 
has been given by Fomalont {\em et al.} (1991) and
Windhorst {\em et al.} (1993) and are summarized below.

(1) After deep cleaning, the contribution from the thermal noise of the radio 
	receivers on the images is
   	almost pure Gaussian fluctuations.  
	 The parameter errors associated with such noise is
     well determined from the Gaussian fitting algorithm for the
     stronger sources, but biases are introduced for the weaker
     sources.  Windhorst {\em et al.} (1984) and Condon (1997) have shown that these routines
     systematically overestimate flux density and angular size due to
     noise bias. To reduce the effect of this bias, for weak radio 
	 sources with a peak flux density less than 6.5~$\sigma$, 
	 we used several different methods to determine 
     flux density and generally assumed that a source
     was unresolved unless it was broadened by more than 50~percent of
     the beam size.  The total error estimate for the sky
     flux density is the quadrature sum of the errors from the 
     image noise, primary beam correction, and antenna
     pointing. The minimum flux
     density error for sources weaker than 6.5~$\sigma$ was set to 25\%.
	 The
     position errors from the Gaussian fits, on the other hand, are
     accurately determined by the algorithm even for the weaker
     sources.

(2) The effects of confusing weak sources are the major problem in
     obtaining reliable fits for the weaker sources.
     Monte Carlo simulations show
     that at $9~\mu$Jy and $3.5\arcsec$
     resolution, sources at the 1 $\mu$Jy level will significantly
     affect roughly 20\% of discrete radio sources
     in the absence of clustering (Richards 1996). The amplitude
     and scale of clustering for the microjansky population is
     uncertain, but may increase confusion by as much as a factor of two
     at this resolution. No correction can be made
     for this effect without higher resolution
     observations. Thus, some of the weaker sources which appear
     resolved may be contaminated by confusion from a weak,
     nearby source. For this reason we have assumed that the sources
     between 9 and 15 $\mu$Jy, which appear slightly resolved in the
     fitting procedure, are contaminated by an unrelated background
     source. We therefore list them as unresolved with conservative
     angular size limits. 
    
(3) We corrected the peak flux densities and angular sizes for
     chromatic aberration following Bridle \& Schwab (1989).
     The finite sampling of the visibility data produced insignificant
     smearing in all the images.
  
(4) We estimate the errors in the primary beam correction following
     Windhorst {\em et al.} (1993). This uncertainty increases
     monotonically from about three percent near the phase center to
     14\% at the eight percent power point (Napier 1992).  For sources
     beyond this
     point we only give estimates of the flux density which we believe 
	 are accurate to
     within a factor of two.

\subsection{The Radio Source List}

	The 41 sources found within 8 arcmin from the field
center and with a peak flux density greater than or equal to 9.0~$\mu$Jy
are listed in Table 3. Of
these, 29 sources lie within the 8\% attenuation contour of the
primary beam (within a 4.6 arcmin radius of the phase center) and form
a statistically complete sample to the 99\% confidence level.  We have
listed the radio sources in order of increasing Right Ascension.  The
nomenclature follows pseudo-IAU convention with truncated angular
position.

	Eighteen of the sources listed in Table 3 were also included in
our earlier paper based on the C and CnB array data alone (Fomalont {\em
et~al.}  1997). For sources far away from the phase center of the
radio image (i.e., in the flanking fields), the calculated sky flux
densities were overestimated by Fomalont {\em et al.}  by up to a
factor of two due to an error in the primary beam correction.
However, other parameters for these sources agree with these
new observations within the estimated errors.

A description of Table 3 is as follows.  All uncertainties are given
at the one sigma level.

{\em Column} (1) --- The source name. Sources with an asterisk are 
	             {\em not} in the complete sample.

{\em Column} (2) --- The Right Ascension (--12 hr.) in J2000
 		     coordinates.
	
{\em Column} (3) --- The uncertainty in Right Ascension. 

{\em Column} (4) --- The Declination (--62$\deg$) in J2000
		     coordinates.

{\em Column} (5) --- The uncertainty in Declination.

{\em Column} (6) --- The peak flux density (S$_p$) in microjansky as
	             measured on the
		      3.5\arcsec \hskip5pt image, unless otherwise noted.

{\em Column} (7) --- The integrated flux density (S$_i$) in microjansky.

{\em Column} (8) --- The {\em deconvolved} (FWHM) Gaussian source size,
	             $\Theta$, is given in arcsec.
 		      For unresolved sources upper limits 
		      are shown.

{\em Column} (9) --- The position angle of the major axis on the sky 
		     measured in degrees.

{\em Column} (10) --- The integrated sky flux density (S$_{8.5 GHz}$)
                       after correction for the primary-beam
    		       attenuation.

{\em Column} (11) --- The uncertainty in the integrated sky flux density.

\section{THE OPTICAL COUNTERPARTS}
	
	The Hubble Deep Field is the most sensitive optical image of
the sky yet obtained. In the 4.7 square arcmin region, objects as
faint as U = 27.6, B~=~28.1, V~=~28.7, and I = 28.0
(AB magnitudes) are detected at the 10 $\sigma$ level. Williams {\em
et al.} (1996) describe the WFPC2 observations and data reduction
procedure and present a complete catalog of objects detected within
the HDF.  In addition to the HDF, eight HST exposures of one 
orbit were taken in I--band (F814) in areas immediately adjacent to
the HDF (Williams {\em et al.} 1996).  The point source sensitivity
for each of these frames is about I~=~25.

        We also used a deep Palomar 200-inch R-band frame taken by
Steidel (1996a) for additional optical identifications and astrometric
use.  The limiting magnitude in this image is about R = 27 with a resolution
of about $1.5"$. This field is nine arcmin on a side and
covers somewhat more area than the HFFs.

     From previous observations of microjansky radio fields on deep
Palomar 200 inch Four-shooter mosaics complete down to R~=~27 and
HST WFPC mosaics complete to R $\simeq$~26, the identification
fraction for faint radio sources is typically about 90\%, with the
reliability of the identification sample generally
exceeding 95\% (Windhorst {\em et al.} 1995,
Fomalont {\em et al.} 1991).  Thus, we expect the optical completeness
of the microjansky sources in this survey of the HDF to approach
100\%.

\subsection{Radio/Optical Alignments}

	In order to make reliable optical identifications of the radio
sources, we must align the HST and VLA coordinate grids. Although the
fine guidance system of the HST is accurate to a few milliarcsec,
intrinsic offsets in the HST Guide Star Catalog positions on the order of
1--2 arcsec are the limiting source of error in tying the HST
astrometric grid to any global coordinate system.  The coordinate
system we used at the VLA is linked to that of the FK5/J2000 quasar
inertial reference frame through the calibrator source 1217+585.  In
order to reliably identify the radio sources with their optical
counterparts, we need to bring the HDF and its flanking fields onto
this grid.

    The position grid of the VLA HDF images are tied to the radio source
1217+585 with an assumed position of $\alpha$ = 12\hh 17\mm 11.0202\ss
\hskip5pt and $\delta$~=~58$^\circ$35\arcmin 26.228\arcsec \hskip5pt
(J2000) (Patnaik et al 1992).  This position is known to be within
0.013\arcsec~of the FK5 reference frame.  The  accuracy of transfering
the position of 1217+585 to the position of the VLA images is less
than 0.05\arcsec, derived from an estimate of typical long term
systematic phase errors and the angular separation in the sky between
1217+585 and the Hubble field. Thus, the radio position reference
frame for the radio data presented in this paper is within 0.1\arcsec
~of the FK5 reference frame.  The position uncertainties associated with
each radio source are in all cases larger than the above radio
astrometric uncertainty.

    The most accurate method of aligning a radio image with an optical
image is by using the `obvious' identifications of many small diameter
radio sources with galaxies.  Quite often both the radio and optical
cores of quasars and AGN have small angular sizes and can be superimposed
to an accuracy of
less than 0.1\arcsec. This comparison has already been made between
the brighter radio sources in the Hubble Deep Field and the offset
between the HDF and the quasar inertial reference frame is less than
0.05\arcsec~(Williams {\em et al.} 1996).

     Because of the small number of radio source identifications on
each of the eight Hubble flanking fields, their alignment with
the radio reference frame cannot be made with sufficient accuracy.
For this reason we used a two step process in registering the Hubble
flanking frames to the VLA radio frame.  First, we determined possible
identifications for 25 radio sources on the deep
Palomar 200-inch optical frame (Steidel 1996a).  We then
determined the offset, the overall plate scale and plate rotation
needed to minimize the radio-optical displacement of the objects.  The
resulting residual offset of a typical radio-optical coincidence
was $0.3\arcsec\times 0.2$\arcsec~in Right Ascension and Declination, 
using the best 19 identifications.
We thus believe that the registration between the
Palomar image and the VLA image is better than 0.1\arcsec~in each
coordinate.  No large-scale distortions were apparent in the Palomar
image.

    Next, we aligned the eight Hubble flanking fields to the
Palomar image by using six good quality coincidences of symmetric and
relatively bright galaxies for each of the 24 Hubble frames (three
CCDs per flanking field).  Only a Right Ascension and 
Declination displacement was determined
since each Hubble flanking field covers a small area and the a priori
pixel separation and orientation in the sky are well known. 
Offsets as large as two arcsec were found between the Palomar
frame and individual flanking fields, with typical uncertainties
of about 0.1\arcsec. We estimate that the final alignment between the 
radio and optical grids is better than 0.2\arcsec.

\subsection{Radio Identifications: The Nine Microjansky Sample}

         In Figure 2 (Plate 1) we show the radio contour image
overlayed onto the Hubble Deep Field image.  We made identifications
of radio sources in Table 3 with optical objects in the HDF,
the flanking fields, and the Palomar image.  Using the alignment
of the VLA radio image with these optical images,
we can evaluate, with some confidence, any significant displacement
of a tentative radio and optical identification in terms of the
likelihood of a chance coincidence.

     The probability of a chance identification depends on the
magnitude of the optical object and the radio-optical separation.  The
probabilities reported in Table 4 are derived from an analysis on the
Palomar field in the following manner: 1) The radio image was shifted
randomly with respect to the Palomar image within the range of 10 arcsec to
30 arcsec with random angle.  2) Using a series of search radii between
0.2 arcsec and 3 arcsec, the probability of coincidence with an optical
object as a function of magnitude was determined.  3) This simulation
was repeated 10,000 times in order to determine the probability distribution
of chance identifications.  By using the actual galaxy
distribution and radio source distribution, these probabilities should
reflect the density and clustering present in the
Hubble field region in both the radio and optical images.

	The above analysis gives an estimate of the likelihood
of a chance radio-optical galaxy alignment. Galaxy counterparts
to all the sources in the complete radio sample were found
by searching for galaxies within a three arcsec radius
from the radio centroid. The galaxy with the highest
identification reliability was taken to be the real 
optical counterpart. The reliability of the seven sources 
in the complete radio catalog within the HDF are all better
than 99 percent with radio-optical offsets of order 0.1\arcsec--0.2\arcsec ,
consistent with the errors expected from the positional
uncertainties.
	
	We also matched all radio sources detected 
in regions covered by the HFFs and the Palomar image
using the same criteria. This yielded nineteen additional identifications.
Three radio sources 
had no reliable identification to I~=~25 on the HFF images and to
R~=~27 on the Palomar images.  Thus the identification fraction for this
survey is about 26/29~=~90\% to R~=~27, comparable to that found for 
previous microjansky surveys.

\subsection{Radio Identifications: The Supplementary Sample}

     With the accurate alignment of the radio and optical coordinate
grids, we are able to identify some brighter galaxies with radio
sources which are below the completeness limit of $9.0~\mu$Jy.  For
example, within the HDF there are 17 positive peaks between 3.5--5.0
$\sigma$ as opposed to 11 {\em negative} peaks less than 3.5 $\sigma$.  
The excess number of
positive peaks is due to faint sources contained in the radio image
but below the formal completeness limit. However, without additional
information, we cannot determine which of the positive peaks (about
one third) correspond to real sources in the sky.

     Since we find that the virtually all of the sources stronger than
9 $\mu$Jy are identified with bright galaxies, we expect the
slightly weaker sources to be similarly identified.  Hence, a near
coincidence of a weaker source between 6.3 and 9.0 $\mu$Jy with a
relatively bright galaxy will confirm with high probability that
the weak radio source is real.

	Our search criterion was to look for any galaxy 
brighter than R = 25 located within two arcsec 
(equal to about twice the radio positional error)
of the 3.5--5.0 $\sigma$ radio sources within 4.6\arcmin~
of the field center. These 19 additional radio sources, 
seven which are located in the HDF, are shown in Table 5.
All of these supplementary sources were too weak to derive angular sizes,
so only peak image and sky corrected flux densities are given.

--TABLE 5 Here (The supplementary Radio Source Identification List)--

    In order to provide a check on the statistical reliability of this
additional sample, we performed the same analysis on the 11 {\em
negative} radio peaks found in the HDF at less than --3.5
$\sigma$. None of these were located within two arcsec of a
galaxy brighter than R = 25. This is the result expected from
Table~4 which gives the reliability matrix for an identification.
We also measured the flux density for each supplementary
source in a preliminary 1.4 GHz image made of the field
(Richards 1998). Positive flux density was found for 
each source, providing further evidence to their reality.
Thus, we believe that essentially all of
the sources listed in Table 5 are real and that we have properly
identified their optical counterparts.

\subsection{Catalog of Hubble Deep and Flanking Field Identifications}

      Table 6 lists identifications for
all of the sources listed in Tables 3 which are part of the
complete radio sample and the supplementary sources listed in Table 5.
Table 6 is
divided into three parts: a) the Hubble Deep Field, b) the HFFs and c) the
Palomar field and is arranged as follows:

{\em Column} (1) ---  The source name. Sources denoted with an '\dag ' are
                      taken from the supplementary sample.

{\em Column} (2) ---  The optical identification in Williams {\it et al.} 1996. 

{\em Column} (3) ---  The radio--optical positional offset in arcsec.
	
{\em Column} (4) ---  The integrated sky flux density (S$_{8.5GHz}$) in microjanskys.

{\em Column} (5) ---  The integrated optical R-band
	              magnitudes (AB) as provided by Steidel (1996a).

{\em Column} (6) ---  Average near infrared flux density taken in the H+K passbands
                      bands (Cowie {\em et al.} 1997).

{\em Column} (7) ---  ISO mid-infrared flux density as measured
		      by Goldschmidt 
		      {\em et al.} (1997). Sources with 
		      flux densities greater than 100 $\mu$Jy
	              are at 15 $\mu$m while those less
                      than 100 $\mu$Jy are at 6.7 $\mu$m.
	
{\em Column} (8) --- Spectroscopic redshift as measured by the
Keck/HDF consortium (Cohen {\em et al.} 1996, 1998,
Moustakis {\em et al.} 1997, Phillips {\em
et al.} 1997, Steidel {\em et al.} 1996b).

% Redshifts in parenthesis are taken from the photometric catalog
% of either Mann {\em et al.} (1997) or Lanzetta, Yahil \&
% Fernandez-Soto (1996).

{\em Column} (9) ---  The logarithm of the monochromatic luminosity of
                      the radio source given in W/Hz. These values are
                      calculated assuming H$_o$ = 50~km/s/Mpc, q$_o$ =
                      0.5. The 1.4-8.5 GHz spectral index where
                      known is used to 
                      correct for the K-correction in individual
                      sources (Richards 1998).
                      For those sources without
                      a measured spectral index, a value of 
                      $\alpha = 0.35$ is assumed (Windhorst {\it et
                      al.} 1993).
 		      
{\em Column} (10) ---  Galaxy type based on color, light profile,
	              and visual morphology. E~=~elliptical,
                      Sp = spiral, Irr/Mrg = Irregular/Merger, 
		      U~=~uncertain

{\em Column} (11) ---  Reliability of the optical identification.

{\em Column} (12) --- Contour levels used in Figure 3 in microjanskys/beam.

-----TABLE 6 Here (Radio Source Identifications in the Hubble Fields) ---

\section{DESCRIPTION OF THE IDENTIFICATIONS}

     One of the goals of our investigation is to study the optical 
	 counterparts of radio sources in the Hubble Deep Field and the
surrounding fields.  The discussion here will only summarize the
properties of the identifications (and empty fields). Further
analysis on the optical properties of the radio sources in this field
is in progress (Windhorst {\em et al.} 1998).

        We have produced overlays of the optical and radio images for
all of the sources in Table 6.  These overlays are shown in
Figure 3 and are presented in the same order as 
in Table~6.
The contours show the radio emission and the grey-scale
 the optical emission.  Most of the optical frames are from the
Hubble Deep Field (V band) or the Hubble Flanking Fields (I band)
 which have a resolution better than 0.2\arcsec. Five of the outlying
radio sources are not
covered by one of the HFFs; these sources are compared with the
1.5\arcsec ~resolution Palomar image. The optical contrast was chosen
to best show the radio and optical emission and is
not meant to be used in any quantitative manner. Each of these
individual images is approximately 15\arcsec ~on a side
and oriented with north to the top and east to the left.

	We now consider some of the properties of individual sources
including redshifts, intrinsic luminosity, spectral features
(if known), unusual colors, sizes, morphology, and emission at 
other wavelengths.  Information for several sources is also available
in the the near infrared (Cowie {\em et al.} 1997, 
Hogg {\em et al.} 1997, Dickinson {\em et al.} 1998) 
in the  J, H, and K bands, and
in the mid-infrared at 7~$\mu$m and 15 $\mu$m (Goldschmidt {\em et
al.} 1997).  
The optical spectral information was provided
by the Keck/HDF consortium (Cohen {\em et al.} 1996, 1998,
Moustakis {\em et al.} 1997, Phillips {\em
et al.} 1997, Steidel {\em et al.} 1996b).

	We assign preliminary optical morphological types
(sprial, irregular, merger, and/or elliptical) from
examination of the indvidual optical identifications.
More detailed classification based on colors 
and surface brightness profiles is in progress.

	For those sources with reliable distance indications,
we calculate star-formation rates (SFR) based on the radio flux
density measurements. We only consider those sources with
L$_{8.5GHz} < 10^{23.5}$ W/Hz assuming that all sources more
luminous are powered (at least in part) by an AGN (e.g., Condon
1989). It
is of course quite possible that the lower luminosity radio
sources, as well, contain AGN. We
used the conversion of thermal radio emission to SFRs given
by Condon (1992). We assume that each radio source has the same
ratio of thermal to non-thermal radiation as observed in
M82 (Carlstrom \& Kronberg 1991) and calculate SFRs according 
to the K-corrected fraction of thermal emission observed in
each galaxy. These estimates correspond only to SFRs for
stars more massive than 5~\Msun, as stars less massive
contribute little to either the synchrotyron or thermal
radio luminosity. In order to convert our value to integrated star
formation rates between 0.1 to 100~\Msun for a Salpeter
initial mass function, these rates should be multiplied by 200.

 	For a check on our star formation rates derived from 
the radio luminsoity, we transformed
the SFRs as calculated by Rowan-Robinson {\em et~al.} (1997) based
on ISO measurements to those for massive stars alone using a Salpeter
initial mass function.  
In calculating SFRs we assume
no contamination of the radio emission from an AGN. As the opacity 
of galaxies at all radio frequencies of interest greater than 8.5 GHz
is negligible, our measurement are upper limits on the true SFRs.
Likewise the absence of radio emission from any galaxy in the Hubble
field allows us to place a firm upper limit on its star-formation
rate. At redshifts of 0.2, 0.5, 1, and 2 these limits correspond
to SFRs of about 1, 10, 70, and 600 \Msun /yr, respectively for any
given galaxy.

	For a few sources, we provide radio spectral indices, $\alpha$,
based on preliminary 1.4 GHz flux density measurements made with the
VLA in its A-array during Novemember 1996 (Richards 1998), using the
convention $S_{\nu} \propto \nu ^{-\alpha}$.

	Sources marked with an '\dag ' are taken from 
the supplementary radio sample.

\subsection{Optical Counterparts in the HDF}

{\bf 3641+1142}: This disrupted group of galaxies is almost certainly
part of a merging system. There are at least six distinct optical
components brighter than I = 22 within a three arcsec radius of the
radio emission peak. At a $z$ = 0.585, the corresponding linear
separation is about 20 kpc and the absolute magnitude,
M$_B$ = --21. The radio emission, if due to star-formation
alone, corresponds to a SFR of about 59 \Msun /yr.  This
compares to the estimate given by Rowan-Robinson {\em et al.}  (1997)
of about 22 \Msun /yr from a 7 $\mu$m flux density of 52
$\mu$Jy. 
%Interestingly, the radio spectral index is rather flat, with
%$\alpha <$ 0.4, which may indicate the presence of substantial thermal
%emission.

{\bf 3644+1249}:   The identification is with the galaxy to the
north which lies at a redshift of 0.557. The object
located about two arcsec to the south is at {\em z}~=~0.555,
corresponding to a physical separation of about 15 kpc if they are
at the same recessional velocity. This possibly merging system
has several prominent emission lines including [OII], [OIII], and H$\beta$
with P Cygni profiles (Cowie {\em et al.} 1997).
The galaxy is rather blue (B--V = 0.6)
and has an H+K magnitude of 19.0 (Cowie {\em et al.} 1997).
It is also listed in the ISO catalog of Goldschmidt {\em et al.} (1997)
with $S_{15 \mu m}$ = 319 $\mu$Jy. With $L_{8.5 GHz}$ = $10^{22.5}$
W/Hz, this system has an estimated star-formation rate of 
31 \Msun /yr. Rowan-Robinson {\it et al.} give a SFR $\sim$ 17 \Msun /yr.
The radio spectral index is $\alpha$ = 0.7 $\pm$ 0.2.

{\bf 3644+1133}: The optical identification is with a bright (M$_B$ = 
--23) elliptical galaxy at $z$ = 1.013.  The core of the radio source
is unresolved ($\theta < 0.1\arcsec$) and has a flat radio spectrum
($\alpha$ = 0.1 $\pm$ 0.1) with steep spectrum emission oriented N-S
and extending about 15\arcsec.  This source is about 10\%
linearly polarized in the 3.5\arcsec \hskip5pt image (P. A. =
180\deg), consistent with its classification as an FR-I
radio source. Its radio luminosity is about $10^{24.8}$ W/Hz.
	
	This interesting radio source has been detected by ISO with
$S_{7 \mu m}$ = 50 $\mu$Jy.  The mid-IR radiation may be due to 
the stellar continuum as suggested by Rowan-Robinson {\em et 
al.} (1997).

	The half light radius of a de Vaucouleurs surface brightness
fit to the I-band image is about 1.8\arcsec, or about 16 kpc at this
redshift, large even for a giant elliptical galaxy. Another 
morphologically disturbed radio source (3641+1142) is located
20 arcsec away and is at a similar redshift. There are also at
least four other galaxies within 1.5 arcmin of the central elliptical
with $z$ = 1.013 $\pm$ 0.005.

	We also note the presence of an elongated chain of galaxies,
or possibly a merging system in the northern radio lobe of 3644+1133 and
at a similar redshift ({\em z} = 1.016).  This feature is similar in
appearance to the 'chain' galaxies described by Cowie {\em et al.}
(1995), although it appears much larger in extent than the galaxies
discussed by those authors. It is not clear what relation
exists, if any, between 3644+1133 and the 'chain' galaxy which has
multiple, apparently unresolved 'hot spots.' One possibility is that
the 'chain' galaxy is undergoing star-formation induced by the
plasma jet emerging from the red elliptical.
 
{\bf 3646+1404}:	A nearly face on spiral at $z$ = 0.960 is 
coincident with the radio core (size less than 0.1\arcsec). We
observed this source to vary by about 30\% over an 18 month period.
Variability in addition to a broad emission line spectrum 
(Phillips {\em et al.} 1997) and an inverted radio spectral
index of $\alpha = -0.1 \pm 0.1$, suggests that           
this  spiral galaxy contains an AGN. It also has a red color of V -- I = 1.8.

	This radio source was detected by ISO with $S_{7\mu m}$ = 
52 $\mu$Jy. Although Rowan--Robinson {\em et al.} (1997) interpret this source 
as a massive starburst
with a SFR $\sim$ 200 \Msun /yr,
star-formation seems unlikely to be powering the bulk of the
radio emission in this variable source. 
	
{\bf 3646+1226}\dag :  Although {\em not} in our complete radio sample,
this 4.7 $\sigma$ radio source 
is also seen at the 5 $\sigma$ level on our 1.4 GHz map.
However, there is no obvious optical counterpart
to this radio source to the HDF survey limit of R~=~29. 
If real, it is unlikely that this unidentified radio source
is an intrinsically faint galaxy at moderate redshift as no other
radio sources are known to be identified with such faint galaxies
(at a redshift of one the host galaxy would have to be 
fainter than about $M_V$ = --14 to escape detection). The
observed radio emission could be the displaced lobe of an extended
source with asymmetric structure and no detectable radio emission
from the parent galaxy, but we consider this also unlikely as
other microjansky sources all appear coincident with their optical
counterparts to within a few arcsec. More likely, it may be
a very high redshift ({\em z} $>$ 6)
{\em R-band dropout} galaxy, in which case near infrared observations
with HST/NICMOS may be able to detect the host galaxy.

{\bf 3647+1255}\dag : This identification is with a reddish (V -- I = 1.2) 
irregular galaxy at {\em z}~=~2.931. This system may be merging as it 
appears highly asymmetric on the HST frame.  It has the highest observed 
redshift of any object in our sample.

{\bf 3648+1416}\dag :   This radio emitter is a spheroidal
 galaxy at $z$ = 2.008 and is included in the Lyman break 
catalog of Steidel
{\em et al.} (1996b) as C2--05. Active star-formation
is suggested by the weak emission lines with P Cygni profiles.
The radio emission may be caused by a combination of 
star-formation
and an AGN with a total power of $L_{8.5} = 10^{24.0}$~W/Hz. 
As pointed out by Steidel {\em et al.} a companion 
galaxy located three arcsec to the north is likely
at the same redshift 

{\bf 3648+1427}\dag :   The optical counterpart to the radio emission is a
bright elliptical located on the edge of the HDF.
% At a redshift
%f 0.023 (Mann {\em et al.} 1997), the  absolute
%magnitude is about M$_{I}$ = --17,
%haracteristic of a dwarf elliptical.
 The optical colors are rather
blue with V -- I = 0.3. Steep spectrum radio emission may be
associated with the source ($\alpha = 0.7 \pm 0.3$).
%nly a very modest star-formation rate of 0.02 \Msun /yr
%s needed to explain the radio emission (P$_{8.5GHz}$ = 19.3 W/Hz), 
%lthough the presence of a very weak AGN cannot be ruled out.
The 15 $\mu$m flux density (S$_{15 \mu m}$ = 231 $\mu$Jy) is consistent
with thermal re-radiation of AGN light from a hot dusty torus (Rowan-
Robinson {\em et al.} 1997).

{\bf 3649+1313}:  The identification in this crowded
optical field is with a spiral galaxy at ${\em z}
= 0.475$. Low level radio emission may also be detected
from a fainter spiral galaxy located about 3 arcsec 
to the north. The brighter galaxy is also in the ISO catalog
of HDF sources with $S_{7 \mu m}$ = 42 $\mu$Jy. Our estimate of
SFR = 29 \Msun/yr is in reasonable agreement with that given by 
Rowan-Robinson
{\em et al.} of SFR $\sim$ 19 \Msun/yr. The radio
source has a steep spectrum with $\alpha = 0.8 \pm 0.2$.
	
{\bf 3651+1321}\dag : This radio source is associated with a face-on spiral
galaxy at {\em z} = 0.199.
The optical spectrum shows significant H$\beta$ emission. The implied
star-formation rate, as calculated from the radio flux density, 
is about 1~\Msun per year. The galaxy may also a source of significant
mid-infrared emission with a S$_{15 \mu m}$ = 472 $\mu$Jy.

{\bf 3651+1221}: We identify this radio source with the bright R = 21.8 magnitude
spiral galaxy located 0.8\arcsec ~to the south of the radio centroid. 
However,
diffuse optical emission is present directly underneath
the centroid of the radio contours which may
be either a separate low surface brightness galaxy 
or an indication of tidal disruption of the fainter
optical galaxy to the north.
        The spiral is at a redshift of
0.299 and has L$_{8.5 GHz}$ = 10$^{21.9}$ W/Hz. It
is quite red (V -- I = 1.3) and has an H+K
magnitude of 18.5. The presence of strong [OII] emission
(Cowie {\em et al.} 1997) could be indicative of either
star-formation or nuclear activity. If the steep spectrum
($\alpha$ = 0.8 $\pm$ 0.2 ) radio emission is
associated with star formation, a rate of 10 \Msun /yr
is inferred.
 
{\bf 3651+1226}\dag :  The optical identification is a compact
system at unknown redshift. It has a surprisingly red optical-
infrared colors, (H+K) - R = 5, implying possibly a high redshift
or early type galaxy counterpart.

{\bf 3652+1354}\dag :  The galaxy counterpart is possibly either a 
chain galaxy or a multiple merger, as it shows evidence for six distinct
optical nuclei with blue colors (B -- V = 0.2) within about 3 arcsec 
of the radio emission. The optical
spectrum is dominated by very strong [OII] emission and places the
galaxy at a redshift of 1.355. If the radio emission is not contaminated
by an AGN component, then we infer a SFR $\sim$ 200 \Msun /yr.
It is similar in optical morphology to 3641+1142.

{\bf 3655+1311}: This radio source is an elliptical galaxy at
$z$ = 0.370 as determined from an absorption line spectrum.
This implies a $L_{8.5 GHz} = 10^{22.1}$~W/Hz.  The galaxy may 
harbor a weak AGN as it has a rather flat radio spectral 
index ($\alpha <$ 0.3). The companion galaxy located about two
arcsec to the west is at an estimated 
 redshift of 0.36 (Lanzetta, Yahill \& Fernandez-Soto
1996) and may be triggering the AGN activity.

{\bf 3656+1302}:  This steep spectrum radio source 
($\alpha = 1.0 \pm 0.3$) is identified with
 an elliptical galaxy.
% at an estimated redshift of 1.215 (Lanzetta,
%hil \& Fernandez-Soto 1996). 
A $S_{7 \mu m}$
= 37 $\mu$Jy object in the ISO catalog is located five arcsec  
to the northwest and may be associated with this galaxy. 

\subsection{Optical Counterparts in the HFFs}

{\bf 3632+1105}: The optical identification is with a
relatively bright (R~=~20.1) spiral galaxy. The radio spectral index
is flat ($\alpha <$ 0.3).

{\bf 3633+1431}\dag :  A spiral galaxy located at {\em z} =
0.519 is identified with this radio source. It may be part 
of a larger group including 3634+1434 and 3635+1435.
We calculate a SFR~=~27~\Msun /yr for this L$_*$ galaxy 
($M_V$ = --20.7).

{\bf 3634+1434}\dag : This bright spiral galaxy may be 
part of an interacting group including 3633+1431 and 3635+1435.

{\bf 3634+1212}: This bright disk galaxy ($M_V$ = --22.7)
has a steep radio
spectral index ($\alpha$ = 0.7 $\pm$ 0.1) and is also detected by ISO
with $S_{15\mu m} = 726~\mu$Jy. The optical image shows evidence for a 
double nucleus which may indicate a recent merger. The radio emission
is consistent with starburst activity with an estimated SFR = 100
\Msun /yr.

{\bf 3634+1305}\dag :   This weak radio source is associated
with a spiral galaxy. 

{\bf 3634+1240}: This radio sources is at a redshift of 1.219 and has
a monochromatic luminosity of $L_{8.5 GHz}$ = $10^{24.3}$ W/Hz.
With a spectral index of $\alpha$ = 0.7 $\pm$ 0.1 and a mid-IR 
flux density as detected by ISO of 444 $\mu$Jy, this source is
likely to harbor
substantial active star-formation. However, the inferred 
SFR $\sim$ 1000 
\Msun /yr may indicate the presence of an embedded AGN.
The optical counterpart
is a Scd galaxy.

{\bf 3635+1435}:   The optical counterpart to this 
steep spectrum radio source ($\alpha$~=~0.8~$\pm$~0.3)
is a low surface brightness galaxy, possibly an 
irregular. It may also be associated with the group including
3634+1435 and 3633+1431.

{\bf 3635+1424}:   The optical identification is with
a face on spiral galaxy. The radio spectral index
suggests optically thin synchrotron emission with
$\alpha = 0.9 \pm 0.3$.

 {\bf 3637+1135}: At a redshift of 0.078, this 
  source has a radio luminosity of
 $L_{8.5 GHz}~=~10^{20.7}$~W/Hz, typical of
 normal spiral systems. This galaxy was also
 detected by ISO with 
 $S_{15 \mu m}$~=~420 $ \mu$Jy. The radio emission is
 steep spectrum radio emission ($\alpha$~=~0.6~$\pm$~0.2) and 
  likely associated with disk star-formation with an 
 estimated rate of 0.4~\Msun /yr. 
  
{\bf 3638+1116+}: This apparent disk system may be interacting
with neighboring galaxies located about two
arcsec to the northwest and southeast. The radio
spectral index is $\alpha$~=~1.1~$\pm$~0.3.

{\bf 3639+1249}: This steep spectrum radio source
($\alpha$ = 1.0 $\pm$ 0.3) is coincident with a 
spiral galaxy. The radio extension may be real or due
to weak source confusion and it is also apparent on
the 1.4 GHz image. The galaxy is a bright ISO source
with S$_{15 \mu m}$ = 433 $\mu$Jy.

{\bf 3640+1010}: We tentatively identify this 
radio source with a very compact
I = 25 magnitude galaxy. However, there is no counterpart
visible on the Palomar frame to R = 27, possibly
due to beam dilution or a spectral break in the
optical continuum.  The 
radio source has a spectral index
of $\alpha$ = 0.6 $\pm$ 0.2.

{\bf 3641+1130}\dag: We identify this weak radio source
with the bright spiral galaxy to the northeast despite
the two arcsec offset. Confusion from the galaxy 
to the southwest may be present. The galaxy 
lies at {\em z} = 0.089 and possibly has a 
star-formation rate of 0.3 \Msun /yr. It is 
also an ISO source with S$_{15 \mu m} = 420~\mu$Jy.

{\bf 3642+1331}: This compact ($\theta < 0.1 \arcsec $)
steep spectrum ($\alpha$ = 0.9 $\pm$ 0.1) 
radio source has a very red optical counterpart
(R -- I $>$ 2) at the detection threshold in the HST
I band image. Near infrared observations by
Hogg {\em et al.} (1997) and Dickinson
{\em et al.} (1997) cover the area which show a 
K = 21.2 magnitude galaxy at the location of the identification
but with no detectable emission in J, yielding 
a color of J -- K  $>$ 2.3. The radio-optical positional
offset of about 0.5\arcsec ~is likely real as it is greater than
both the positional and registration error combined. Still
the chance probability of finding an I = 25 magnitude galaxy within
0.5\arcsec ~of this radio source is only about 1\%. 

	The peak position of the K band source agrees with that of
3642+1331 to better than 0.2\arcsec ~and the likelihood
of finding a K = 21.2 magnitude galaxy within this
error circle is much less than 1\% based on the 
known surface density of infrared objects in the HDF
(Hogg {\em et al.} 1997);
thus we believe this to be the likely identification.
The optical counterpart is possibly either heavily dust obscured
or a radio galaxy at a substantial redshift. 

{\bf 3642+1305}+: The radio emission is coincident with a
peculiar system, possibly disrupted by the  
spiral to the south. The 6.7 $\mu$m flux density
is S$_{7 \mu m} = 51 \mu$Jy.

{\bf 3642+1502}\dag : The optical identification is with a spiral galaxy.

{\bf 3646+1448}:  This radio source is clearly resolved
from neighboring 3646+1447 in higher resolution 
1.4 GHz and 8.5 GHz images. Although an R = 23.9 magnitude galaxy lies
about 3\arcsec ~to the north and has a 20\% chance 
coincidence probability, the displacement is about 15 times the
radio positional error. Thus we classify this radio source as 
unidentified down to the best limits of R = 27. 
The radio source has a spectral index of $\alpha$ = 0.8 $\pm$ 0.2.

{\bf 3646+1447}: The optical identification is with a compact,
possibly elliptical galaxy. The radio spectral index is
a steep $\alpha$ = 0.7 $\pm$ 0.2.

{\bf 3651+1030}: This disk galaxy lies at a redshift of 0.410
   and has a radio luminosity of $L_{8.5 GHz} = 10^{22.6}$ W/Hz. 
   If the radio luminosity is dominated by star-formation 
   then a rate of 33 \Msun /yr is inferred. The radio 
   spectral index is 0.6~$\pm$~0.1.

{\bf 3652+1444}: At a redshift of 0.322, this radio source
has a modest luminosity of $L_{8.5 GHz} = 10^{23.0}$ W/Hz and
an inverted spectral index, $\alpha$~=~--0.5 $\pm$~0.1. This source is
observed to vary in intensity on the time scale of months,
suggesting the presence of an AGN.
	
{\bf 3653+1139}: An R = 22.3 magnitude compact spiral galaxy is 
associated with this radio source. It has a fairly steep
spectral index,  $\alpha$ = 0.9 $\pm$ 0.1 indicating
the presence of optically thin synchrotron emission which is
somewhat extended in the 1.4 GHz image. It is also detected 
at 15 $\mu$m with a flux density of 138 $\mu$Jy.

{\bf 3658+1434}\dag :   This radio source is identified 
with a face-on spiral at a redshift of
0.678. The implied
star-formation rate is 52 \Msun /yr.  

{\bf 3701+1146}: This extended radio source is possibly
   identified with an R = 22.4 magnitude spiral galaxy located 
   nearly three arcsec to the south. The likelihood of 
   a chance coincidence between this radio source and
   a galaxy of this magnitude is about 5\%. However, the
   radio-optical positional offset is too large
   to be attributed to positional uncertainties. The radio spectral index
   is  $\alpha$ = 0.5 $\pm$ 0.1. If the source
   is associated with a spiral galaxy, then it is 
   unlikely to be a one sided radio lobe; thus the nature of this
   radio source remains unclear.

{\bf 3707+1408}: Although optically unidentified 
  (R $>$ 27), this radio source is detected in both
   the 8.5 GHz and 1.4 GHz image. It has an intermediate 
   spectral index of $\alpha$~=~0.4~$\pm$~0.2. The radio emission
   is likely extended as it is only detected in the 10\arcsec ~resolution
   8.5 GHz image. This radio source is very unlikely to 
   be associated with a typical disk galaxy if we assume the
   radio emission associated with such sources is confined
   to the galactic disk. This would place a redshift limit
   of about 0.2 from angular size arguments, assuming
   disks are no larger than about 30 kpc. This is inconsistent
   with the magnitude limit unless the system has about eight
   magnitudes of extinction in the visual. 
	
	One possibility is that the source is a displaced
radio lobe as discussed for 3646+1226.
Alternatively, the source may be a distant
   radio galaxy. Assuming microjansky radio sources are
   not associated with galaxies fainter than $M_V = -17$ (the
   faintest extragalactic radio counterpart observed), we can
   place a lower limit on redshift of the optical counterpart of about 1.
	
{\bf 3708+1056}: A moderately bright (R = 20.4)
  edge-on spiral galaxy is identified with this
  radio source and is coincident with the optical
  nucleus. The radio emission has a flat spectral 
  index, $\alpha$ = 0.4 $\pm$ 0.2.

{\bf 3708+1246}: Identified with a faint disk galaxy,
  this radio source may be partially confused with a galaxy to the 
  northeast. The absence of observed radio emission
  at 1.4 GHz places a limit on its spectral index of 
   $\alpha <$ 0.3.

{\bf 3711+1054}\dag :   The optical identification is a spiral 
   galaxy.

{\bf 3721+1129}:  This radio source has an inverted spectrum ($\alpha$
= --0.4 $\pm$ 0.2) and may be associated with a distorted
disk galaxy. However, the radio-optical offset of
over one arcsec is too high to be accounted 
for by astrometric errors which is the dominant
uncertainty in comparing the radio and optical 
emission for this source (note the good agreement
positional agreement for 3725+1128 which is 
on the same HFF frame).

	If the irregular galaxy to the north is
a foreground galaxy, then the radio source is 
unidentified to R = 27 (Steidel 1996a). Inverted spectrum radio 
sources are generally associated with quasars, luminous
ellipticals, or Seyferts and thus optically
bright.  The nature of this source remains unclear. Intriguingly,
Cowie reports the presence of a faint K band object much closer
to the radio position.

{\bf 3725+1128}:   This slightly extended radio source
is clearly resolved into two components in higher
resolution 1.4 GHz images, with a separation of 
about one arcsec. The optical identification is with 
a compact, possibly elliptical, galaxy located between the two radio lobes.
The radio spectrum is steep with $\alpha = 1.0 \pm 0.1$,
indicating optically thin
synchrotron emission.

\subsection{Optical Counterparts in the Palomar Image}

{\bf 3642+1545}:  The radio emission is coincident with 
a R = 22 magnitude galaxy of uncertain classification. It has
a moderately steep spectral index of $\alpha$~=~0.6 $\pm$~0.1.

{\bf 3657+1455}:   This possibly resolved radio source is 
identified with a R = 22.8 magnitude galaxy and also has 
$S_{15 \mu m}$ = 210 $\mu$Jy. This source is located near
the edge of a HFF frame but no optical emission
is apparent in the HFF image. The radio spectral index 
is $\alpha$~=~0.7~$\pm$ 0.2.

{\bf 3700+0908}: There is no optical counterpart to the
limit R = 27 for this radio source.  
The radio emission is fairly steep spectrum with
$\alpha$ = 0.9 $\pm$ 0.1. The optical counterpart
might possibly 
escape detection if it is very compact 
and hence diluted on the 1.5 arcsec Palomar image.
 
{\bf 3711+1331}: Steep spectrum ($\alpha$ = 0.8 $\pm$ 0.2),
extended radio emission is coincident with an R = 23.1
galaxy of unknown classification. There is
some evidence on the optical frame for interaction
with a companion galaxy. 

{\bf 3716+1512}: A relatively bright (R = 20.4) possibly
merging system is identified with this slightly
resolved radio source. The radio spectral index is 
$\alpha = 0.2 \pm 0.1$. 

\section{DISCUSSION OF THE HUBBLE FIELD RADIO GALAXIES}

\subsection{The Optical Counterparts to the Microjansky Sources}

%We consider the magnitude distribution of the microjansky 
%dentifications in this field in comparison with the results of
%indhorst {\em et al.} (1995) and Hammer {\em et al.} (1995), who
%onducted deep searches for optical counterparts of microjansky 
%adio sources.  These
%wo samples are nearly all identified and contain 29 and 17
%ources respectively with V and I band photometry. This allows for an
%ccurate estimate of the R magnitudes of these sources for comparison
%ith the R magnitudes of the HDF and HFF radio sources as provided by
%teidel (1996a).  Histograms of the magnitude distribution of the
%dentifications in the present survey and the surveys of Hammer {\em
%t al.} (1995) and Windhorst {\em et al.} (1995) are shown in Figure
%.  Both distributions are similar with a peak near R = 22.5.
%he agreement amongst the various fields is excellent, indicating
%hat we are observing representative samples of the microjansky
%opulation.

	Spectroscopic redshifts are available for a
large sub-sample of the microjansky sources with optical
identifications. The 18 currently available redshifts for the 
HDF and HFF radio are shown in Figure 5.
The mean redshift is about $z \sim$ 0.8, although identifications
range from 0.1 to nearly 3.

%ources are compared with those from the surveys of Hammer {\em et
%l.} (1995) and Windhorst {\em et al.} (1995) in Figure 5.  
%he mean redshift is about $z \sim$ 0.8, although identifications
%ange from 0.1 to nearly 3.

	We use the available redshifts to calculate the monochromatic
radio luminosity. Figure 6
shows their distribution.
The typical identified radio source has a radio luminosity of about 
$10^{23}$ W/Hz.

\subsection{The Nature of the Emission From the Microjansky Sources}

      We consider three interpretations of the microjansky radio
emission from galaxies found in this survey.

1) The observed radiation is caused by a central engine (AGN).  In this
case the bulk of the radio emission is confined to
the nuclear region
of the galaxy. These sources are characterized by their 
small angular dimensions on the order of a milliarcsec. This can 
lead to a significant
opacity to synchrotron self-absorbtion and flat
or inverted radio spectra typically in the range
$-0.5$\lsim$\alpha$\lsim 0.5. These radio sources are usually associated
with luminous elliptical galaxies, Seyfert galaxies, and quasars.

2) The observed radiation is synchrotron emission due to the
acceleration of electrons in supernova remnants.
These sources are optically thin and have radio spectra which
reflect the energy distribution of the relativistic electrons,
usually within the range $0.5$\lsim$\alpha$\lsim 1. Flatter 
radio spectra have been observed in some very compact 
($\theta <1\arcsec$) nuclear starbursts where the principal
contribution to the radio emission is believed to 
be synchrotron emission (Condon {\em et al.} 1991). These
sources are almost exclusively associated with disk
galaxies (i.e., irregulars, mergers, and spirals).

3) The observed radiation is free-free emission from 
hot gas associated with star-formation in HII regions concentrated
along the galactic disk and bulge. These sources have optically
thin thermal spectra with $\alpha \sim -0.1$ and are 
identified with disk galaxies.

	First, we consider the optical morphology of the
microjansky counterparts. Of the 41 identifications with
HST imaging, 30 are disk galaxies (spirals, irregulars, 
and/or mergers), comprising over 70\% of the sample.
We classify seven microjansky counterparts as 
ellipticals (17\%), although this fraction could be
as high as 30\% as the morphology of
several identifications remains uncertain
or unknown. This result is consistent with the fraction 
found by Windhorst {\em et al.} (1995), although
it appears to be somewhat lower than the fraction
found by Hammer (1996). 
For the purposes of this discussion, we assume all elliptical
galaxy counterparts contain an AGN.

	Next, we consider the radio morphology. Extended
radio sources with P $< 10^{25}$~W/Hz are not 
likely to be FR II radio galaxies, so their radio emission 
is plausibly caused by 
star-formation. We have accurate angular size information
for only 19 radio sources in the complete sample of which 
12 are resolved (58\%). Two of these sources, 3644+1133
and 3725+1128 are suspected AGN based on other arguments
(see notes in \S 6) while three remain unidentified. The remaining 
seven sources are likely associated with disk star-formation.

	Another important clue is provided by the 
radio spectra. We assume 
sources with radio spectral index less than $\alpha$ = 
--0.2 are synchrotron self-absorbed and thus 
are associated with an AGN.
Three radio sources in this sample can be classified
as AGN with this criterion and we place an
upper limit of six such inverted sources in the complete 
sample ($<$ 21\%).

	On the basis of the above analysis, we 
identify 11 probable low luminosity AGN in this survey (24\% of the
sample): 3634+1240, 3644+1133,
 3646+1404, 3646+1447,
3648+1424, 3648+1416, 3651+1226, 3652+1444, 3655+1311, 3656+1302, and 3725+1128.

\subsection{Microjansky Radio Emission from the Disk Galaxies}

	Radio emission from disk galaxies can result from
either star-formation 
or from AGN activity connected with a central engine. In the previous section we have
used the observed angular sizes of a sub-sample of this
survey to argue that their emission likely comes from 
extended star-formation regions. The average spectral index for
these sources is about $\alpha~\sim$~0.4, substantially flatter
than that observed in local extended spirals ($\alpha \sim$ 0.8). 
Thus it is possible that we are observing significant fractions
of thermal emission in these systems. High frequency observations 
might resolve this question. 
 
	But what of the other 22 disk systems in our sample? 
At least one of these, 3646+1404, contains an AGN. Ten of 
these 22 galaxies are detected by ISO
at mid-infrared wavelengths including 3646+1404
(Goldschmidt {\em et al.} 1997).
Of these, five appear to be in interacting groups
or merging systems. However, the mid-infrared radiation 
could be  UV light from an embedded AGN or
massive O and B stars reprocessed by dust. At the angular resolution and 
sensitivity available in this VLA survey, we are currently unable to
to discriminate between AGN activity and star-formation
for the majority of disk galaxies.

\section{CONCLUSIONS}
	
	We have catalogued a total of 60 radio sources (29 in a
statistically complete sample with flux density greater 
than 9.0 $\mu$Jy) in a deep survey at 8.5 GHz in a field
16 arcmin in diameter and centered on the Hubble Deep Field. Of these
sources, 14 lie within the HDF itself, of which seven are in our
complete sample. Thirty-four sources are covered by
the HST flanking fields and ground based optical observations, of
which three remain unidentified to R = 27. An additional
radio source detected in both the 8.5 GHz and 1.4 GHz image
within the HDF remains unidentified to R = 29 and may be
a very high redshift object. The optical data indicate
that microjansky radio galaxies are composed primarily of spiral
and irregular/merging systems (70-90\%), with a lesser
fraction (10-30\%) of elliptical galaxies. Although the
strong radio luminosity/density evolution observed at 
microjansky levels is dominated by disk galaxies, it
is unclear to what degree star-formation and/or low
luminosity AGN play a role.
	
	Microjansky radio sources are more distant and optically more
luminous than average field galaxies and are
distributed over a wide range of redshift space from 0.1 to 3. 
None of the
sources in the sample presented in this paper are identified with
either galactic stars or quasars. Unlike the sources found in radio
surveys conducted at higher flux density, the sources here are
primarily located in nearly "normal" host galaxies, whose
bolometric luminosity is dominated by starlight rather
than an AGN.

	A complete optically identified sample of microjansky sources
with redshift information will allow a more precise discussion of the
evolution of different types of faint radio sources.  A
quantitative understanding of this evolution will provide important
information on the time scale for the formation of AGN  and their duration,
and on the star-formation history of field galaxies, which is tightly
linked to radio emission at 8.5 GHz. 

\vskip20pt

	Many individuals kindly made their data available to us,
including the HDF team, the Keck/HDF consortium which provided
many redshifts, and particularly L.~Cowie who made his K-band data
and redshifts available to us.  We especially thank C. Steidel who 
provided us with
his Palomar images and spectroscopic information in advance of
publication. We also acknowledge the assistance C. Burg and S. Odewahn 
with the VLA/HDF astrometry. We are thankful to J. Condon
for his insightful comments and suggestions during the course 
of this work. R. Mann commented on an earlier version of this 
manuscript.

E. A. R. acknowledges the
support of a Sigma Xi Grant-in-Aid-of-Research.  Support for part of
this work was provided by NASA through grant AR-6337.*-96A from the
Space Telescope Science Institute, which is operated by the Association
of Universities for Research in Astronomy, Inc., under NASA contract
NAS5-2655, and by NSF grant AST 93-20049.  Haverford and Bryn Mawr
students John Eppley, Daria Halkides and Andrew West contributed to
the analysis of the VLA data; their participation was supported in
part by a grant from the William Keck Foundation to the Keck Northeast
Astronomy Consortium.  This work has made use of observations made at
the Kitt Peak National Observatory, National Optical Astronomy
Observatories, which is operated by the Association of Universities
for Research in Astronomy, Inc. (AURA) under cooperative agreement
with the National Science Foundation.

\newpage

\section{REFERENCES}

\nref Benn, C. R., Rowan-Robinson, M., McMahon, R. G., Broadhurst, T. J.
      \& Lawrence, A. 1993, MNRAS, 263, 123

\nref Bridle, A. H. \& Schwab, F. R. 1989, in  {\em Synthesis
      Imaging in Radio Astronomy}, PASP conf. series, 6, 259

\nref Carlstrom, J. E. \& Kronberg, P. P. 1991, ApJ, 366, 422

\nref Cohen, J. G. {\em et al.} 1996, ApJL, 471, 5

\nref Cohen, J. G. 1998, in {\em Proceedings of the STScI symposium on the 
Hubble Deep Field}, eds. M. Livio \& M. Donahue (Cambridge: Cambridge Univ. Press), in press 

\nref Condon, J. J. 1989, ApJ, 338, 13

\nref Condon, J. J. 1992, ARA \& A, 30, 575

\nref Condon, J. J. 1997, PASP, 109, 166

\nref Condon, J. J., Huang, Z. P., Yin, Q. F. \& Thuan, T. X. 1991, 
        ApJ, 378, 65

\nref Cowie L. L. {\em et al.} 1997, {\em private communication}

\nref Cowie L. L., Hu, E. M. \& Songaila, A. 1995, AJ, 110, 1576

\nref Dickinson, M. {\em et al.} 1998, {\em in preparation}

\nref Fomalont, E. B. 1996, in {\em Proceedings of Symposium 175: Extragalactic Radio Sources},
eds.
      R. Ekers, C. Fanti, \& L. Padrielli, (Dordrecht: Kluer), 555

\nref Fomalont, E. B., Kellermann, K. I., Richards, E. A., Windhorst, R.
A. \&
  Partridge, R. B. 1997, ApJL, 475, 5

\nref Fomalont, E. B., Windhorst, R. A., Kristian, J. A. \& Kellermann,
K. I.
      1991, AJ, 102, 1258

\nref Goldschmidt, P. {\em et al.} 1997, MNRAS, 289, 465

\nref Hammer, F. 1996, in Science with the Hubble Space Telescope -- II, eds.
      P. Benvenuiti, F. D. Machetto \& E. J. Schreier, 101

\nref Hammer, F., Crampton, D., Lilly, S. J., LeFevre, O. \& Kenet, T.
1995,
  MNRAS, 276, 1085

\nref Helou, G., Soifer, B. T. \& Rowan-Robinson, M. 1985, ApJ, 298, 7

%\nref Holtzman, J. A., Burrows, C. J., Casertano, S., Hester, J. J., 
%     Trauger, J. T., Watson, A. M. \& Worthey, G. 1995, PASP, 107, 1065

\nref Kellermann, K. I., Fomalont, E. B., Richards, E. A., Partridge, R.
B. \&
      Windhorst, R. A. 1998, {\em in preparation}

\nref Kron, R. G., Koo, D. C. \& Windhorst, R. A. 1985, A \& A, 146, 38

\nref Lanzetta, K. M., Yahil, A. \& Fernandez-Soto, A. 1996, Nature,
381, 759

\nref Mann, R. G. {\em et al.} 1997, MNRAS, 289, 482

%\nref Minkowski, 1960, ApJ, 132, 908

\nref Moustakas, L. {\em et al.} 1997,
http://astro.berkeley.edu/davisgrp/HDF/

\nref Muxlow, T. 1997, {\em private communication}

%\nref Napier {\em et al.} 1983, IEEE Proceedings, vol. 71, Nov. 1983,
%     p. 1295-132

\nref Napier, P. 1992, {\em private communication}

%\nref Napier, P. J., Thompson, A. R. \& Ekers, R. D. 1983, IEEEP, 71,
%1295

%\nref Odewahn, S. A. 1995, PASP, 107, 770

%\nref Odewahn, S. A., Windhorst, R. A., Driver, S. P., \& Keel, 
%     W. C. 1996, ApJL, 472, 130

%nref Oort, J. A. \& Windhorst, R. A. 1985, A \& Ap, 145, 405

\nref Patnaik, A. R. Browne, I., Wilkinson, P. N. \& Wrobel, J., 1992, MNRAS,
      254, 655

\nref Phillips, A. C. {\em et al.} 1997, ApJ, 489, 543

\nref Richards, E. A. 1996, in IAU 175: Extragalctic Radio Sources, eds.,
      Ekers, R., Fanti, C. \& Padrielli, L., 593

\nref Richards, E. A. 1998, {\em in preparation} 

\nref Rowan-Robinson, M. {\em et al.} 1997, MNRAS, 289, 490

%\nref Rupen, M. 1997, VLA Test Memoranda, 202 (NRAO)

\nref Steidel, C. C. 1996a, {\em private communication}

%nref Steidel, C. C., Giavalisco, M., Pettini, M., Dickinson, M. \&
%     Adelberger, K. L. 1996, ApJL, 426, 17   
 
\nref Steidel, C. C., Giavalisco, M., Dickinson, M. \& Adelberger, K. L.
       1996b, AJ, 112, 352

\nref Taylor, G. {\em et al.} 1996, The VLA Calibrator Manual (NRAO)

\nref Thuan, T. X. \& Condon, J. J. 1987, ApJL, 322, 9

\nref Williams, R. E. {\em et al.} 1996, AJ, 112, 1335

\nref Windhorst, R. A., van Heerde, G., \& Katgert, P. 1984, A\&A Sup. 58, 1

\nref Windhorst, R. A, Fomalont, E. B., Kellermann, K. I., Partridge, R.
B.,
      Richards, E. A., Franklin, B. E., Pascarelle, S. M.
       \& Griffiths, R. E. 1995,
      Nature, 375, 471

\nref Windhorst, R. A., Fomalont, E. B., Partridge, R. B. \& Lowenthal,
J. D.
      1993, ApJ, 405, 498

%\nref Windhorst, R. A., Kron, R. G. \&  Koo, D. C. 1984, A \& AS, 58, 39

\nref Windhorst, R. A, Miley, G. K., Owen, F. N., Kron, R. G. \&
      Koo, D. C. 1985, ApJ, 289, 494

\nref Windhorst, R. A. {\em et al.} 1998, {\em in preparation}

%\nref Windhorst, R. A., Van Heerde, G. M. \& Katgert, P., 1984,  
%     A \& A, 140, 220

\newpage

\section{FIGURE CAPTIONS}

1. The contour map of the 8.5 GHz image at 3.5 arcsec 
    resolution. Contours are drawn at the
    --5.0, 5.0, 10, 20, 40 and 80 sigma levels
    ($\sigma$ = 1.80 $\mu$Jy).

2. Color image of the HDF with radio contours overlaid. Contour
   levels are at the 3.5~$\sigma$ and 5~$\sigma$ level.

3. Greyscale images of the individual optical identifications.
    Each postage stamp is approximately 15 \arcsec ~
    on a side with north at the top and east to the left.
    The contour levels correspond to the 8.5 GHz emission
    as measured in the 3.5\arcsec ~image as shown in Table 6.
	The order follows that of Table 6.

4. The magnitude histogram for the 42 galaxies in the HDF and HFF
   with R $<$ 27 is shown.
   
5. Histograms for the galaxies with available redshifts are shown 
   for the same samples.

6. Histograms of the radio monochromatic power are shown for
   the same sample.

\bn

\end{document}